\documentclass[10pt,preprint]{aastex}
\shorttitle{[Fe~{\sc ii}] Emission Around P Cygni}
\shortauthors{Smith \& Hartigan}
\begin{document}

\title{INFRARED [Fe II] EMISSION FROM P CYGNI'S NEBULA: ATOMIC DATA,
MASS, KINEMATICS, AND THE 1600~A.D.\ OUTBURST}

\author{Nathan Smith\altaffilmark{1,2,3}}
\affil{Center for Astrophysics and Space Astronomy, University of
Colorado, 389 UCB, Boulder, CO 80309}

%\and

\author{Patrick Hartigan} 
\affil{Department of Physics and Astronomy, Rice University, 6100
South Main, Houston, TX 77005-1892}

\altaffiltext{1}{Hubble Fellow; nathans@casa.colorado.edu}

\altaffiltext{2}{Visiting astronomer at the IRTF, operated by the
  University of Hawaii under contract with NASA.}

\altaffiltext{3}{Visiting astronomer, Kitt Peak National Observatory,
  National Optical Astronomy Observatories, operated by the
  Association of Universities for Research in Astronomy, Inc., under
  cooperative agreement with the National Science Foundation.}

\begin{abstract}

We present moderate and high-dispersion 1--2.5~$\micron$ spectra of
the $\sim$10\arcsec-radius nebula around P Cygni, dominated by bright
emission lines of [Fe~{\sc ii}].  Observed [Fe~{\sc ii}] line ratios
disagree with theoretical transition rates in the literature, so we
use the spectrum of P Cygni's nebula to constrain the atomic data for
low-lying levels of [Fe~{\sc ii}].  Of particular interest is the
ratio [Fe~{\sc ii}] $\lambda$12567/$\lambda$16435, often used as a
reddening indicator, for which we empirically derive an intrinsic
value of 1.49, which is 10--40\% higher than previous estimates.
High-dispersion spectra of [Fe~{\sc ii}] $\lambda$16435 constrain the
geometry, detailed structure, and kinematics of P Cygni's nebula,
which is the major product of P Cygni's outburst in 1600 A.D.  We use
the [N~{\sc ii}]/[N~{\sc i}] line ratio to conclude that the nebula is
mostly ionized with a total mass of $\sim$0.1 M$_{\odot}$; more than
the mass lost by the stellar wind since the eruption.  For this mass,
we would expect a larger infrared excess than observed.  We propose
that the dust which obscured the star after the outburst has since
been largely destroyed, releasing Fe into the gas phase to produce the
bright [Fe~{\sc ii}] emission.  The kinetic energy of this shell is
$\sim$10$^{46.3}$ ergs, far less than the kinetic energy released
during the giant eruption of $\eta$ Car in the 1840s, but close to the
value for $\eta$~Car's smaller 1890 outburst.  In this respect, it is
interesting that the infrared spectrum of P Cygni's nebula resembles
that of the ``Little Homunculus'' around $\eta$ Car, ejected in that
star's 1890 eruption.  The mass and kinetic energy in the nebulae of
$\eta$ Car and P Cygni give insight to the range of parameters
expected for extragalactic $\eta$ Car-like eruptions.

\end{abstract}


\section{INTRODUCTION}

P~Cygni is a key object for understanding mass loss in the most
luminous evolved stars.  At a distance of 1.7 kpc (adopted throughout
this paper; Najarro et al.\ 1997) it is the nearest luminous blue
variable (LBV), and it is one of only two stars in our Galaxy to have
been observed during a giant LBV eruption (P~Cyg's eruption in 1600
and $\eta$~Car's giant eruption in the 1840s; see Humphreys et
al. 1999).  Giant eruptions of LBVs --- where the bolometric
luminosity increases by a few magnitudes and the progenitor star
survives the event --- have been recognized in other galaxies as well;
examples include the ``Type V'' supernovae SN1961v in NGC~1058 (van
Dyk et al.\ 2002; Filippenko et al.\ 1995; Goodrich et al.\ 1989) and
SN1954j (V12) in NGC~2403 (Tamann \& Sandage 1968; Smith et al.\
2001b; van Dyk et al.\ 2005), and several more recent $\eta$~Car
analogs classified as Type IIn supernovae (e.g., Filippenko 2005; van
Dyk et al.\ 2000; Wagner et al.\ 2004; Drissen et al.\ 2001).  The
lightcurves of $\eta$~Car and P~Cyg showed dips of several magnitudes
after their eruptions (Humphreys et al.\ 1999; de Groot 1988; Lamers
\& de Groot 1992), usually attributed to dust formation.  Nebulae
around these stars provide useful clues to the physics of their
outbursts, such as the explosion geometry, chemical composition,
ejection speed, mass, and kinetic energy.  Since the extragalactic
$\eta$~Car analogs are too distant to spatially resolve their nebulae,
and since $\eta$~Car itself is such an extreme example, P~Cyg's nebula
provides an important benchmark for understanding the evolution of
intermediate-luminosity LBVs (see Smith et al.\ 2004).

P~Cyg's nebula is harder to observe than the Homunculus of $\eta$~Car
because of its weaker contrast with the central star at all
wavelengths.  This difference is partially due to the longer time
elapsed since P~Cyg's eruption (400 yr, as opposed to only 160 for
$\eta$ Car), allowing the ejecta more time to disperse and fade.
Excess infrared (IR) emission from dust in P~Cyg's nebula is much less
pronounced than in $\eta$ Car --- dust emission at 2--20 $\micron$ has
not been detected above the thermal-bremsstrahlung emission from
P~Cyg's stellar wind (Smith et al.\ 2001a; Waters \& Wesselius 1986).
The cool dust excess at far-IR wavelengths (Waters \& Wesselius 1986)
is also meager compared to other LBV nebulae like the Homunculus of
$\eta$ Car (Morris et al.\ 1999; Smith et al. 2003) or the ring nebula
around AG Car (McGregor et al.\ 1988).  While the most basic physical
quantities (mass, kinetic energy) of the ejecta from $\eta$ Car's
giant eruption are now fairly well constrained (Smith et al.\ 2003),
those same quantities for P~Cyg's nebula are not.

P~Cyg is surrounded by complex emission-line nebulosity, including an
inner shell with a radius of $\sim$10--11\arcsec, an outer shell with
a radius of $\sim$1$\farcm$6, and some more extended infundibular
debris in a ``giant lobe'' several arcminutes across (Barlow et al.\
1994; O'Connor et al.\ 1998; Meaburn 2001; Meaburn et al.\ 1996, 1999,
2004).  In this paper we focus on the near-IR spectrum of the bright
``inner shell'', which is also observed in radio free-free emission
(Skinner et al.\ 1998).  Johnson et al.\ (1992) found the inner nebula
to be nitrogen rich, indicating CNO-processed ejecta like in $\eta$
Car (e.g., Smith \& Morse 2004).

A valuable spectral diagnostic of P Cyg's nebula is its bright
emission from IR lines of [Fe~{\sc ii}], such as $\lambda$12567,
$\lambda$16435, and related transitions between the $^4D$ term to the
$^4F$ and $^6D$ terms.  These are collisionally-excited lines with an
upper level $\sim$1 eV above the ground state.  They are bright in the
nebula of P~Cyg and in the nebulae of several other LBVs (Smith 2002a,
2002b, 2005; Hamann et al.\ 1994; Allen et al.\ 1985).  While bright
IR emission from these [Fe~{\sc ii}] transitions is seen in
shock-excited sources like Herbig-Haro objects and supernova remnants
(e.g., Hartigan et al.\ 2004b; Graham et al.\ 1987; Oliva et al.\
1990; McKee et al.\ 1984), and so is often taken as a signpost of
shocks, the excitation mechanism in all LBVs is not uniform.  In some
LBV nebulae the [Fe~{\sc ii}] emission is due to shocks, while in
others UV excitation is responsible (see Smith 2002a).  Analysis of
these [Fe~{\sc ii}] lines is complicated by disagreement in the
literature on their relative transition rates (Nussbaumer \& Storey
1988; Quinet et al.\ 1996; Bautista \& Pradhan 1998).

Preliminary near-IR spectra of P~Cyg in the H band showed bright
[Fe~{\sc ii}] $\lambda$16435 out to 10\arcsec\ from the star (Smith
2001, 2002a), consistent with the inner shell seen in images (Barlow
et al.\ 1994).  Here we investigate the IR emission from [Fe~{\sc ii}]
in more detail with spectral observations from NASA's Infrared
Telscope Facility (IRTF), sampling many [Fe~{\sc ii}] transitions
across the 1--2.5 $\micron$ wavelength range at moderate resolution
using Spex, and at high spectral resolution for the 16435 \AA\ line
alone using CSHELL.

\section{OBSERVATIONS}

\subsection{SPEX}

We obtained long-slit 1-2.5 $\micron$ spectra of P Cygni's nebula on
2004 August 19 with the near-IR spectrograph Spex (Rayner et al.\
2003) mounted on the IRTF.  Spex uses a 1024$\times$1024 Aladdin3 InSb
array, with a spatial pixel scale of 0$\farcs$15.  We used Spex in
long-slit mode with the 0$\farcs$3-wide slit (providing a spectral
resolution of $R\approx$2000) oriented along P.A.=270$\arcdeg$, and
positioned 4\arcsec\ north of the star to sample emission from the
nebula while avoiding the bright central star (see Figure 1).  Sky
subtraction was accomplished by nodding 30\arcsec\ along the slit so
that the nebula was always in part of the slit; the total on-source
exposure time was 16 minutes in each of four filters (K, H, and two
filters in the J band).  Wavelength calibration was measured using an
internal emission lamp, while flux calibration and correction for
telluric absorption were accomplished using observations of
spectroscopic standard stars.  On the night these observations were
obtained, the conditions were photometric and the seeing was
$\sim$0$\farcs$8.

Figure 2 shows the spectrum extracted from a 2\arcsec-wide segment of
the long-slit data, corresponding to a bright knot in the nebula
located 4\arcsec\ north and 8\arcsec\ east of the star (see Fig.\ 1).
However, this is not a true sample of the nebular emission, since
there is significant contamination from instrumental scattered
starlight, and starlight reflected by dust (continuum emission and
stellar wind lines are seen).

Therefore, Figure 3 shows the same spectrum, but with the reflected
starlight subtracted.  The reflected starlight was sampled from a
different region along the slit closer to the star where the continuum
emission was much stronger relative to the nebular emission (it was
simple to disentangle the contributions of nebular emission and
scattered light, since [Fe~{\sc ii}] lines like $\lambda$16435 are
absent in the spectrum of the central star).  Intensities of nebular
lines measured in this spectrum are listed in Table 1, where noise in
fitting the continuum level dominates the uncertainties.  Additional
uncertainties in the flux calibration between various wavelength
regions depend on the seeing and the exact position of the slit
aperture during observations of the standard star, potentially causing
changes in the fraction of the total point spread function included in
the slit; these uncertainties are harder to quantify.  However, it is
reassuring that the scattered continuum light from the star in Figure
2 follows a smooth trend from one band to the next --- there is no
strong discontinuity in the continuum level between the $J$ and $H$
bands, for example, indicating no severe difference in slit position
between exposures of the standard star.  Figure 4 shows tracings of
the $\lambda$12567 and $\lambda$16435 intensities and ratios along the
slit, as well as the $\lambda$15335/$\lambda$16435 flux ratio, which
is a diagnostic of variations in the electron density.

\subsection{CSHELL}

On 2003 July 31 we observed P~Cygni with the high-resolution
spectrograph CSHELL (Greene et al.\ 1993) mounted on the IRTF.  These
observations were conducted remotely from the University of Colorado.
CSHELL has a 256$\times$256 SBRC InSb array with a spatial pixel scale
of 0$\farcs$2, although only 160 pixels are illuminated in the spatial
direction, yielding an effective slit length of roughly 30\arcsec.  We
used a slit width of 0$\farcs$5, providing a spectral resolving power
of $R\approx$43,000 or 7 km s$^{-1}$.  The circular variable filter
(CVF) wheel isolates a single order, and we chose a central wavelength
corresponding to the bright [Fe~{\sc ii}] $\lambda$16435 emission line
(vacuum wavelength 16439.98 \AA).

We observed three separate pointings with the slit oriented at
P.A.=0\arcdeg\ and centered on the central star, 3\arcsec\ east, and
4\arcsec\ west (see Fig.\ 1).  At each position, we used total
on-source integration times of 20--30 minutes, and sky-subtraction was
accomplished with identical observations of an off-source position.
The resulting 2-D spectra at these three pointings are shown in Figure
5.  Wavelengths were calibrated using telluric absorption lines,
adopting vacuum wavelengths in the telluric spectrum available from
NOAO.  Velocities in Figure 5 are heliocentric, and have been
corrected for the motion of the Earth (to convert to LSR velocities
for P~Cygni, add 18.1 km s$^{-1}$).

\subsection{Optical Spectra}

To supplement these IR spectra (specifically for the purpose of
investigating the electron density and ionization fraction), we also
obtained visual-wavelength spectra of the nebula.  We observed P Cygni
on 2005 May 5 using the Goldcam spectrograph on the Kitt Peak National
Observatory (KPNO) 2.1m telescope.  We used two different settings to
cover the full optical spectrum from about 3400 -- 7200 \AA.  For blue
wavelengths, we used grating 58 in second order, and for red
wavelengths we used grating 47 in first order.  Both observations used
a 1$\farcs$2$\times$5\arcmin\ slit aperture offset 4\arcsec\ north of
the star (identical to the Spex slit position shown in Fig.\ 1).
Total exposure times were 30 minutes in the blue, and 40 minutes for
the red spectrum.  Flux calibration was performed with reference to
spectra of the central star, P Cygni itself, for which the visual
spectral energy distribution has been studied extensively (e.g.,
Lamers et al.\ 1983).  Some relative line intensities are discussed
further in \S 5.1 and \S 5.2 to aid our interpretation of the IR
spectra.

%% Spex stuff
\section{ANALYSIS OF RELATIVE LINE INTENSITIES}

\subsection{General Comments on the Near-IR Spectrum}

After subtracting the scattered starlight, the near-IR spectrum of
P~Cygni's nebula in Figure 3 is dominated by lines of [Fe~{\sc ii}].
The strength of the [Fe~{\sc ii}] lines is unusual compared to
circumstellar nebulae around other types of evolved stars, such as
planetary nebulae and Wolf-Rayet ring nebulae, and is a testament to
the high density of the gas, and the high luminosity and relatively
low effective temperature of the central star.  The bright [Fe~{\sc
ii}] spectrum of P Cygni's nebula resembles that of the Homunculus
nebula around $\eta$~Car (Smith 2002$b$), although P Cygni shows no
emission lines of molecular hydrogen.  The lack of H$_2$ is probably
due to the much lower mass of P Cygni's nebula (see \S 5.3) and to the
extreme youth of the Homunculus.  The lack of H$_2$ is also consistent
with the fact that we detect no emission from the 2.3 $\micron$ CO
emission features.  These bright near-IR lines of [Fe~{\sc ii}] are
common in the circumstellar nebulae of LBVs in general (Smith
2002$a$).

The long-slit aperture allows us to investigate spatial variations
across the nebula.  For example, Figure 4 shows that across the Spex
slit placed 4\arcsec\ north of the star, there is little variation in
either the reddening or electron density, traced by the [Fe ~{\sc ii}]
ratios $\lambda$12567/$\lambda$16435 and
$\lambda$15335/$\lambda$16435, respectively.  This verifies that in
general, the physical conditions we derive for one particular region
of the nebula probably apply to most of the bright [Fe~{\sc ii}] shell
as well.

Figure 3 and Table 1 also show several emission lines of [Ni~{\sc
ii}].  Of particlar interest are [Ni~{\sc ii}] $\lambda$23079 and
$\lambda$23688 (both are $a^4F-a^2F$), which are rarely seen in
circumstellar nebulae.\footnote{Another line of this multiplet,
[Ni~{\sc ii}] $\lambda$19388 ($a^4F-a^2F$), is expected to be strong
but was not within the wavelength range of our observations.}  These
lines share the same upper term as [Ni~{\sc ii}] $\lambda$7379 and
$\lambda$7412 ($a^2D-a^2F$), which are extremely bright in the nebula
of P Cygni (Johnson et al. 1992; Barlow et al.\ 1994), and whose
strength is thought to be a consequence of fluorescence (Lucy 1995).
These same lines are also strong in the Crab Nebula (Dennefeld \&
Pequignot 1983; Henry et al.\ 1984), and are seen in Herbig-Haro jets
very close to their sources (Hartigan et al.\ 2004a).  It is not known
if a similar continuum-fluorescence process affects the [Fe~{\sc ii}]
lines.

\subsection{Atomic Data}

The numerous [Fe~{\sc ii}] lines in our P~Cyg spectra make it possible
to constrain Einstein A coefficients.  These calculations are
important because the A values for these transitions computed in two
previous studies (Quinet et al.\ 1996, hereafter Q96; Nussbaumer \&
Storey 1988, hereafter NS88) differ by as much as 30\%. A 30\%\ error
in the [Fe~{\sc ii}] $\lambda$12567/$\lambda$16435 flux ratio, used to
estimate reddening because these lines originate from the same upper
level, translates into a much larger error when reddening corrections
are extrapolated to optical wavelengths.  For example, if the
intrinsic ratio of [Fe~{\sc ii}] $\lambda$12567/$\lambda$16435
increases by 30\% , then an object which would have had A$_V$=0 using
the lower ratio now has A$_V$=2.85.

The spectra in Fig.~2 recorded ten [Fe~{\sc ii}] emission lines
between the second-excited term $^4D$ and the first-excited term
$^4$F, and an additional seven [Fe~{\sc ii}] lines from $^4D$ to the
ground term $^6D$. The lowest 13 energy levels of Fe~{\sc ii}, are
respectively, $^6D_{9/2,7/2,5/2,3/2,1/2}$, $^4F_{9/2,7/2,5/2,3/2}$,
and $^4D_{7/2,5/2,3/2,1/2}$, and in what follows we simply refer to
these levels as 1 through 13. For example, in this notation the
12567~\AA\ and 16435~\AA\ transitions are 10-1, and 10-6,
respectively.

Table~2 compiles the 17 observed transitions into 30 line ratios in
four groups, where each group originates from a common upper
state. Within each group we adopted the A value for the brightest
transition from NS88, whose compilation agrees somewhat better with
our data than do the values of Q96. Other A values within the group
then follow from the observed emission line ratio and known reddening
(A$_V$=1.89, Lamers et al.\ 1983).  Table~2 shows the predicted
emission line ratios for A$_V$=0 and A$_V$=1.89 for Q96 and NS88,
together with the line ratios predicted from the new list of A-values
in Table~3, reddened with A$_V$=1.89.  By construction, the new
dereddened A coefficients match the observations in Table~2. The last
column of Table~2 is the uncertainty in the observed ratio.

The first and third lines of Table~3 are of particular interest to all
reddening estimates using near-IR [Fe~{\sc ii}] lines. The new ratio
of A values for the [Fe~{\sc ii}] lines at 12567 \AA\ and 16435 \AA\
is 1.13, $\sim$10\%\ higher than the value of 1.04 of NS88 and over
40\%\ higher than the Q96 value of 0.79.  This corresponds to an
intrinsic [Fe~{\sc ii}] $\lambda$12567/$\lambda$16435 line intensity
ratio of 1.49, higher than Q96 (1.04) or NS88 (1.36).  As Table~3
shows, when NS88 and Q96 disagree most we typically find A values
closer to NS88, but often the new values do not lie between the
estimates of NS88 and Q96.  Thus, our results indicate that previous
studies that adopted the intrinsic $\lambda$12567/$\lambda$16435 line
ratios from Q96 or NS88 have underestimated the true reddening.  The
$\lambda$12567/$\lambda$16435 line ratio is often used to estimate
reddening in AGN as well; Rodriguez-Ardila et al.\ (2004) have also
argued that the A-values of Q96 and NS88 are incorrect, but they favor
an even {\it lower} value for the intrinsic line ratio of only 0.98.
Our study of P Cygni reveals that this cannot be correct, since even
our raw {\it observed} value for this ratio is 1.25.

We did not attempt to constrain collision strengths between the
various levels or compare A values between different groups because
the densities within the P~Cygni nebula are not known well, and the
flow is clumpy (see \S 4). However, such a calculation could be done
with a better understanding of the electron densities in the shell. A
deeper optical search for faint emission lines would help in this
regard.

%%% CSHELL stuff
\section{GEOMETRY AND KINEMATICS}

\subsection{Spatial and Kinematic Morphology}

Figure 5 shows high-dispersion spectra of [Fe~{\sc ii}] $\lambda$16435
at three different positions in the nebula, with the slit oriented in
the north/south direction and centered 3\arcsec\ east of the star, on
the star, and 4\arcsec\ west.  The morphology seen in Figure 5 is
consistent with three slices through a roughly spherical expanding
shell.  However, it is obviously not a uniformly smooth sphere, as
prominent knots and corrugations exist.  The detailed structure of the
shell is reminiscent of Vishniac instabilities when a thin shell with
dense knots is swept up by a faster low-density wind (e.g.,
Garcia-Segura et al.\ 1996).  This is indeed the case, as the shell
expansion is slower than the terminal velocity of the present-day
stellar wind (see \S 4.2).  Aside from these perturbations with
typical sizes of 1--2\arcsec, the only striking departure from
spherical symmetry is a possible blow-out at --80 to 0 km s$^{-1}$
(heliocentric) at the southern edge of the shell, most apparent in
Figure 5$b$.

Although the overall shape of the nebula is basically spherical, it
does show some large-scale brightness asymmetry.  For example, in all
three slit positions in Figure 5, the brightest emission is north of
the nebula's center and on the blueshifted side of the shell.  There
are corresponding local brightness maxima on the opposite side (to the
south and redshifted) of the shell in Figures 5$a$ and $b$.  This
hints at a sort of reflected symmetry consistent with cross sections
through an inclined ring or torus of brighter emission embedded in the
shell, with its northern polar axis tilted away from us by
$\sim$45\arcdeg.

The [Fe~{\sc ii}]-emitting knots in Figure 5 have resolved linewidths
of typically FWHM$\approx$15--20 km s$^{-1}$.  This is obviously too
broad for an iron line to be due to thermal broadening, so it
represents dynamic expansion of the gas or perhaps broadening due to
shocks as the stellar wind interacts with slower nebular material.

\subsection{Physical Parameters of the Shell}

The high-resolution spectra in Figure 5 have clearly resolved the
spatial and kinematic structure of P~Cyg's nebula, allowing us to
quantitatively constrain several of its basic physical parameters.
The slit position centered on the star is the most useful as it
crosses through the middle diameter of the nebula, sampling the
largest spatial extent and the highest velocities.  

Figure 6 shows the same data as in Figure 5$b$ for the slit passing
through the star, but with two ellipses drawn to represent the
idealized inner and outer boundaries of a single spherical shell (with
radii of $R_1$ and $R_2$, respectively).  Because of the knots and
corrugations in the nebula, these ellipses are obviously not perfect
boundaries for the shell, but they are useful for guiding our
interpretation of the nebular structure and for eventually estimating
its mass and kinetic energy in \S 5 below.  Derived parameters for
these two surfaces are discussed below and are summarized in Table~4.

{\it Systemic Velocity}: The centroid velocity of each of the two
ellipses gives the center-of-mass velocity of the nebula.  The dashed
vertical line in each panel of Figures 5 and 6 shows the approximate
centroid velocity of the shell, taken to represent the systemic
velocity of P~Cygni at $v_{\rm sys}$=--38$\pm$5 km s$^{-1}$ (the
uncertainty here is dominated by the irregular structure in the shell,
rather than statistical measurement uncertainty).  Relative to the
local standard of rest, this systemic velocity would be about $v_{\rm
LSR}$=--20$\pm$5 km s$^{-1}$, which is in reasonable agreement with \
--22.6$\pm$1.5 km s$^{-1}$ estimated by Barlow et al.\ (1994) from
optical lines.

{\it Expansion Velocity}: The inner boundary of the shell ($R_1$) has
an expansion velocity of $\pm$120 km s$^{-1}$ with respect to the
systemic velocity, and the outer edge ($R_2$) expands at $\pm$152 km
s$^{-1}$.  These are slower than the present-day terminal speed of
P~Cygni's wind, which is roughly 185 km s$^{-1}$ (Najarro et al.\
1997; Lamers et al. 1996).  Thus, it is reasonable to expect a 65 km
s$^{-1}$ shock at the inside surface of the nebula where the faster
stellar wind catches up.  However, this shock will not profoundly
affect the gas temperature or the relative line intensities in the
nebular spectrum, since radiative energy input from the star exceeds
that of the kinetic energy from the wind by a factor of $\sim$10$^4$.
The mean expansion speed of $\pm$136 km s$^{-1}$ that we measure from
[Fe~{\sc ii}] $\lambda$16435 is close to the value of 140 km s$^{-1}$
measured from optical [N~{\sc ii}] lines by Barlow et al.\ (1994), but
significantly higher than the 110 km s$^{-1}$ expansion they measured
from [Ni~{\sc ii}] lines.  Barlow et al.\ explained the discrepancy
between the apparent expansion speeds of [N~{\sc ii}] and [Ni~{\sc
ii}] by proposing that the [Ni~{\sc ii}] emission arises in slower
dense knots that have been overtaken by the stellar wind, producing
more extended bow shocks around the knots that emit [N~{\sc ii}] and
other nebular lines.  Our new results contradict this conjecture,
since the [Fe~{\sc ii}] and [Ni~{\sc ii}] transitions both share the
same low ionization and high critical densities that motivated the
explanation.  The densest knots should emit [Fe~{\sc ii}] as well.

{\it Shell Radius and Thickness}: The inner and outer radii of the
idealized spherical shell in Figure 6 are $R_1$=7$\farcs$8 or
2.0$\times$10$^{17}$ cm and $R_2$=9$\farcs$7 or 2.5$\times$10$^{17}$
cm.  Some emission is seen to extend beyond these boundaries in the
form of dense knots (interior) or diffuse emission (exterior).  The
overall thickness of the shell is $\Delta R$=1$\farcs$9 or
5$\times$10$^{16}$ cm. Aside from a few redshifted knots located
5--6\arcsec\ south of the star, the shell is hollow (i.e. there is no
younger shell expanding inside this one).

{\it Dynamical Age}: If the ellipses in Figure 6 trace spherical
surfaces of the shell, then at a distance of 1.7 kpc, R/$v$ gives the
dynamical age of each feature assuming linear motion.  For $R_1$ the
dynamical age is 531 yr, and for $R_2$ it is 524 yr; these are in good
agreement with one another considering that the uncertainty is of
order $\pm$5\%.  Of course, the dynamical age reflects linear motion
and does not necessarily signify the true age of the ejecta, since the
shell may have been accelerated by the stellar wind or decelerated by
slower ambient material.  The present-day nebular expansion speed (136
km s$^{-1}$) is in fact lower than the stellar wind speed (185 km
s$^{-1}$), suggesting that the shell has decelerated to $\sim$3/4 of
its initial speed.  Coincidentally, the dynamical age we measure
assuming linear motion is 4/3 of the time since the 1600 outburst. In
other words, by assuming linear motion we overestimate the age by
precisely the same factor that the shell's velocity has apparently
slowed.  This agreement would seem too fortuitous if the bright shell
around P Cygni were not ejected in the 1600~A.D.\ outburst.  Given the
lack of any comparably bright ejecta closer to the
star,\footnote{Barlow et al.\ hinted at the possible existence of a
6\arcsec-radius shell, but we do not see evidence for this smaller
shell in our data.  The envelope close to the star seen by Chesneau et
al.\ (2000) corresponds to the outer parts of the stellar wind, and
cannot be the major result of the 1600 A.D.\ outburst.} {\it we
conclude that the bright} [Fe~{\sc ii}] {\it shell in Figures 5 and 6
was indeed ejected during the 1600 outburst and has been decelerated
by interaction with the ambient medium}.

{\it Filling Factor}: While the overall thickness of the shell derived
above is roughly 1$\farcs$9, it is clear from Figure 5 that the shell
is significantly thinner and corrugated with several knots and loops;
i.e. it has complex substructure.  The [Fe~{\sc ii}] $\lambda$16435
emission is dominated by the densest material in clumps and filaments,
so for estimating the mass in the next section, a representative
density filling factor $f$ is necessary.  Assigning a value for the
filling factor based on the observed structure of the nebula is
subjective, but we can nevertheless make an educated guess, since the
filling factor of the thick shell drawn in Figure 6 is clearly not 1.
A value closer to 0.2 seems more appropriate, but an even lower value
might apply if our spatial resolution is inadequate to resolve the
sizes of features in the nebula.  We provisionally take
$f$=0.2$\pm$0.1.

\section{MASS AND KINETIC ENERGY}

In the previous section, we concluded that the bright [Fe~{\sc ii}]
shell seen in our spectra is indeed the major product of the 1600
outburst of P~Cygni.  Thus, the mass and kinetic energy of this shell
trace the total mass ejected and the mechanical luminosity of the
outburst.  If P~Cygni's nebula can be approximated as a spherical
hollow shell, then a rough estimate of the total gas mass of the
nebula can be expressed as

\begin{equation}
M = \mu m_H \, \frac{n_e}{f_H} f \, \frac{4}{3}\pi (R_2^3 - R_1^3)
\end{equation}

\noindent where $\mu$=2.2 is the mean molecular weight for a neutral
(see \S 5.2) He mass fraction of $Y$=0.55 (Najarro et al.\ 1997);
f$_H$ is the hydrogen ionization fraction, and $f$ is the filling
factor in the shell.  The geometry is constrained (see \S 4.2) from
our high-resolution spectra of [Fe~{\sc ii}] $\lambda$16435, while the
filling factor is only marginally constrained.  This leaves the
electron density and the H ionization fraction as the unknown
quantities that need to be addressed before we can estimate the mass
from equation 1.

\subsection{Electron Density}

Johnson et al.\ (1992) measured an electron density of $n_e$=600
cm$^{-3}$ using the [S~{\sc ii}] $\lambda\lambda$6717,6731 doublet,
although their spectra had low sensitivity.  From the same [S~{\sc
ii}] doublet in higher quality data, Barlow et al.\ (1994) measured
higher densities of 700--2000 cm$^{-3}$ at various positions in the
nebula.  In our new optical spectra, with the slit offset 4\arcsec\
north of the star, the [S~{\sc ii}] ratio indicates a range of
electron densities from about 2,500 to 11,000 cm$^{-3}$.  In Figure 7
we show tracings of the spectrum at three positions along the slit
with the reflected starlight subtracted.

However, much of the mass may potentially reside in dense clumps or
filaments that emit [Fe~{\sc ii}] lines, but where the [S~{\sc ii}]
lines are collisionally de-excited.  Therefore, when deducing the mass
from geometric parameters that apply to [Fe~{\sc ii}] emission
regions, it is more appropriate to estimate the average electron
density in the [Fe~{\sc ii}] emitting gas itself, which tends to favor
the highest densities.  The ratio $\lambda$15335/$\lambda$16435 is a
diagnostic of electron density, and the bottom panel in Figure 4
suggests that this ratio has a fairly uniform value of $\sim$0.15
across the nebula.  Adopting the transition probabilities of
Nussbaumer \& Storey (1988)\footnote{While our spectrum of P Cygni's
nebula suggests some potential problems with the atomic data for
[Fe~{\sc ii}], our new transition rates for these two lines do not
differ from those of Nussbaumer \& Storey by more than 40\%.}, the
$\lambda$15335/$\lambda$16435 ratio of 0.15 would suggest an electron
density of roughly 6000--8000 cm$^{-3}$.  This is in agreement with
some of the denser material traced by the [S~{\sc ii}] ratio in our
spectra. We adopt $n_e$=6000 cm$^{-3}$ as representative for the dense
[Fe~{\sc ii}]-emitting gas, with the caveat that it could be in error
by as much as a factor of 2 in either direction.

\subsection{The Ionization Fraction}

To estimate the true mass of the nebula, we must first deduce the
hydrogen ionization fraction f$_H$ = n(H~{\sc ii})/(n(H~{\sc
i})+n(H{\sc ii})).  We experimented with the spectral synthesis code
CLOUDY (Ferland 1996) to simulate a thick shell with the geometry
described above, illuminated by a central star with T$_{\rm
eff}$=1.9$\times$10$^4$~K and L=6$\times$10$^5$ L$_{\odot}$ (Najarro
et al.\ 1997; Lamers et al.\ 1983, 1996).  We found that He was
neutral and H was almost fully ionized in such a situation.  However,
it is not necessarily safe to assume that P~Cyg's nebula is fully
ionized, since the dense stellar wind extinguishes much of the star's
Lyman continuum radiation if H recombines along the flow.  For
example, Najarro et al.\ (1997) find that in the outermost parts of
the stellar wind, the ionization fraction is $\lesssim$ 0.5.
Returning to the CLOUDY simulations, by varying the amount of escaping
Lyman continuum radiation that is extinguished by the wind we could
produce a wide range of values for the ionization fraction in the
nebula (0.05$\la$f$_H\la$1)\footnote{We note that we could not
reproduce the strength of the near-IR lines of [Fe~{\sc ii}] in any of
these CLOUDY calculations.}.  Furthermore, if much of the Lyman
continuum is indeed extinguished and the nebula would otherwise be
mostly neutral, then we may need to include the effects of shock
heating by the stellar wind that overtakes the shell.  Barlow et al.\
(1994) considered shocks to be important in the observed spectrum of
P~Cygni's nebula. In other words, we need more information to
constrain this problem.

We could, in principle, use the observed ratios of [O~{\sc i}] and
[O~{\sc ii}] lines in the nebula to trace the hydrogen ionization
fraction.  In fact, this was the goal in obtaining our optical
spectra.  Unfortunately, we did not detect [O~{\sc i}] $\lambda$6300
or [O~{\sc ii}] $\lambda$3727 in our spectra of P Cygni's nebula,
perhaps owing to a depleted oxygen abundance or very bright scattered
light from the star at blue wavelengths.  However, [N~{\sc ii}]
$\lambda$6583 is a bright line in the nebula, and our optical spectra
also show weak [N~{\sc i}]$\lambda$5200 at about the 2$\sigma$ level
(in this discussion, [N~{\sc i}]$\lambda$5200 refers to the sum of the
individual doublet lines at 5199\AA\ and 5201\AA ).  The large flux
ratio of the [N~{\sc ii}] to [N~{\sc i}] lines suggests that the gas
is mostly ionized.

The relative flux of [N~{\sc ii}]$\lambda$6583 and [N~{\sc
i}]$\lambda$5200 is determined by three parameters: the electron
temperature, electron density, and the ratio of the densities of
[N~{\sc ii}] and [N~{\sc i}].  The average electron density is
$\sim$6000 cm$^{-3}$, as mentioned above, and the electron temperature
for photoionized gas should be $\sim$10$^4$K. The case of a radiative
shock is somewhat more complicated, as the temperature in the [N~{\sc
ii}]-emitting region is higher than that for [N~{\sc i}]. In this case
we adopt values typical for the cooling zones of a low ionization
shock, where T(N~{\sc ii})$\simeq$10$^4$K and T(N~{\sc
i})$\simeq$6500~K.

To obtain the density ratio n(N~{\sc ii})/n(N~{\sc i}), we performed a
full 5-level atom calculation for N~{\sc i} and N~{\sc ii} using the
latest available atomic data for the Einstein-A values and collision
strengths (Hudson \& Bell 2005; Storey \& Zeippen 2000; Bell et al.\
1995; McLaughlin \& Bell 1993; Baluja \& Zeippen 1988; Zeippen 1987;
Froese Fischer \& Saha 1985; Butler \& Zeippen 1984; Berrington \&
Burke 1981; Le Dourneuf \& Nesbet 1976; Pradhan 1976; Dopita et al.\
1976).  Collisions and radiative decay determine the relative
populations of the energy levels and therefore allow us to infer the
density ratio of N~{\sc ii}/N~{\sc i} from the observed emission line
ratios.  Physically, the value of the density ratio n(N~{\sc
ii})/n(N~{\sc i}) is set by collisional ionization, photoionization,
and charge exchange between N~{\sc i}, N~{\sc ii}, H~{\sc i}, and
H~{\sc ii}.

Using our 3$\sigma$ upper limit for the strength of [N~{\sc i}], we
measure a value for the line ratio [N~{\sc ii}] $\lambda$6583/[N~{\sc
i}] $\lambda$5200 greater than 127 in the shell of P Cygni, with the
slit offset 4\arcsec\ north of the star (if our 2$\sigma$ detection is
to be believed, the same line ratio would be 189).  Using the
2$\sigma$ result for the constant temperature (photoionization) case,
we found nitrogen ionization fractions n(N~{\sc ii})/(n(N~{\sc
i})+n(N~{\sc ii})) of 0.90--0.98 for T$_e$ of 5000-20000 K; the
corresponding 3$\sigma$ limits over this temperature range are $>$0.86
to $>$0.98.  For the shock-like case, the observed [N~{\sc
ii}]/[N~{\sc i}] ratio would imply nitrogen ionization fractions of
$>$ 0.73 and 0.80 for the 3$\sigma$ and 2$\sigma$ measurements,
respectively.  So in all cases, we find that nitrogen is mostly
ionized.  The next question is how to convert the nitrogen ionization
fraction into the hydrogen ionization fraction.  If, as should occur
in the cooling zone of a shock, charge exchange is the dominant
process that ties N~{\sc ii}/N~{\sc i} to H~{\sc ii}/H~{\sc i}, then

\begin{equation}
\frac{n(N II)}{n(N I)} = 4.5 \ {\rm e}^{\frac{-0.94}{ T_{eV} }} 
  \ \frac{n(H II)}{n(H I)} \ .
\end{equation}

\noindent The exponential factor arises from the different ionization
potentials of N (14.53 eV) and H (13.59 eV), while the factor of 4.5
reflects the relative statistical weights of the N~{\sc ii}+H~{\sc i}
and N~{\sc i}+H~{\sc ii} outlet channels.  At $10^4$K, n(N~{\sc
ii})/n(N~{\sc i}) = 1.5 n(H~{\sc ii})/n(H~{\sc i}), so that f$_H$ =
2f$_N$/(3-f$_N$), where f$_H$ is the ionization fraction of H and
f$_N$ is the ionization fraction of N. At $10^4$K, when f$_N$ = 0.9,
we find f$_H$ = 0.86. Hence, whenever charge exchange dominates the
ionization states and N is mostly ionized, so is H.

As the ionization parameter (ratio of density of ionizing photons to
total H density) increases, photoionization becomes increasingly more
important relative to charge exchange in determining the ionization
fractions of H and N.  In P Cygni's nebular shell at the adopted
density, the ionization parameter is roughly 10$^{-5.3}$.  The value
for the hydrogen ionization fraction derived from the observed [N~{\sc
ii}]/[N~{\sc i}] ratio (a few percent neutral) agrees with what we
would expect {\it a priori} from this ionization parameter based on
the Lyman continuum flux escaping the thick wind of P Cygni using an
upper limit of $Q_H\la$10$^{44.4}$ s$^{-1}$ from recent modeling work
that is in preparation by Najarro \& Hillier (Najarro 2005, private
comm.; see also Najarro et al.\ 1997).  In any case, the observed
[N~{\sc ii}]/[N~{\sc i}] ratio implies that P Cygni's nebula is mostly
ionized.  This result removes a potentially large source of
uncertainty in estimating the mass of the outburst.

\subsection{Mass and Mechanical Energy of the 1600~A.D.\ Outburst}

Following the previous subsections, if we take $n_{\rm e}$ to be
$\sim$6000 cm$^{-3}$ and 0.8$\la$f$_H\la$0.95, then equation (1) and
the geometric parameters deduced in \S 4.2 give

\begin{equation}
M \simeq 0.1 \, M_{\odot}
\end{equation}

\noindent for the mass of P Cygni's inner shell nebula.  This is only
an order-of magnitude estimate owing to the factor of $\sim$2
uncertainties in the geometric filling factor and the electron
density.

Is this mass estimate plausible?  We can perform a sanity check by
guessing what we think the nebular mass {\it should} be based on the
density contrast of the nebula and the duration of the outburst.  For
a constant spherical post-eruption wind with
$\dot{M}$=3$\times$10$^{-5}$ M$_{\odot}$ yr$^{-1}$ (Najarro et al.\
1997), the particle density at the radius of the nebula is $\sim$100
cm$^{-3}$.  Thus, the nebula has an overdensity factor of roughly 60
compared to the stellar wind at the same radius, presumably marking a
huge increase in the mass-loss rate during the 1600 A.D.\ eruption.
Based on this factor of 60 overdensity of the nebula, we might expect
the mass-loss rate during eruption to have been
$\sim$1.8$\times$10$^{-3}$~M$_{\odot}$ yr$^{-1}$.  This seems
reasonable, since a mass-loss rate much greater than
10$^{-4}$~M$_{\odot}$ yr$^{-1}$ would be required for a significant
pseudo photosphere to form in the wind (Smith et al.\ 2004), as
expected during an LBV outburst (e.g., Appenzeller 1986).  Comparably
large values of $\dot{M}$ would also be required to form dust in the
wind and obscure the star.  So, if mass loss occurred at this rate
during the 55 yr duration of P Cygni's 1600~A.D.\ eruption (Humphreys
et al.\ 1999), then the nebular mass should be about 0.1~M$_{\odot}$.
This is in fortuitously-good agreement with equation (3).  This
conjecture depends on several assumptions and is not conclusive --- it
merely shows that the mass we derived is sensible.  In particular,
this suggests that our assumed filling factor is not seriously wrong.

Given the high mass-loss rate during the 1600 outburst and the dip in
the historical lightcurve after the eruption, it is reasonable to
assume that significant quantities of dust formed during the event.
Dust residing at the radius of the [Fe~{\sc ii}] shell observed in our
spectra should have a temperature of 60~($Q_{abs}/Q_{em}$)$^{1/4}$~K,
where $Q_{abs}/Q_{em}$ depends on the grain size.  For large grains
like those seen in $\eta$ Car and other LBVs (Smith et al.\ 2003), we
should therefore expect dust temperatures of $\ga$60 K in P Cyg's
nebula, emitting primarily at wavelengths longer than 30 $\micron$.  A
total dust mass $M_d$ emits a flux of

\begin{equation}
F_{\nu} = \frac{M_d \kappa_{\nu} B_{\nu}(T)}{D^2}
\end{equation}

\noindent where $\kappa_{\nu}$ is the grain opacity, $B_{\nu}$ is the
Planck function, and $D$=1.7 kpc.  In the 60 $\micron$ {\it IRAS}
band, for example, 60~K dust with $M_d\simeq$10$^{-3}$~M$_{\odot}$
(assuming a gas:dust mass ratio of 100) and
$\kappa_{\nu}$(60~$\micron$)=87 cm$^2$ g$^{-1}$ (Draine 2003) would
emit a flux of roughly 24 Jy.  Waters \& Wesselius (1986) reported 60
$\micron$ excess emission (flux in excess of the stellar wind
emission, possibly due to cool dust) of roughly 90\%, corresponding to
an excess of only $\sim$1.2 Jy.  This would seem to indicate that
either 1) in our analysis above we have overestimated the mass by more
than an order of magnitude, or 2) that dust grains have been depleted
in P Cygni's nebula by more than an order of magnitude compared to the
typical ISM fraction.  The extreme strength of infrared [Fe~{\sc ii}]
lines that we observe in the nebula suggests that grains are in fact
depleted so that Fe is released into the gas phase.  We found a
similar ionized gas to dust ratio (about 1000) in the nebula of RY
Scuti (Smith et al.\ 2002; Gehrz et al.\ 2001).

If the shell mass is 0.1~M$_{\odot}$ and it is expanding at an average
speed of 136 km s$^{-1}$, then the kinetic energy of P Cygni's shell
is about 2$\times$10$^{46}$ ergs.  This represents the minimum amount
of mechanical energy imparted to the ejecta during the 1600
A.D. eruption.  The luminous energy radiated during that eruption was
roughly 2.5$\times$10$^{48}$ ergs (Humphreys et al. 1999).  Thus, a
small fraction ($\la$1\%) of the total energy was used to accelerate
the ejecta during this event.  Of course, we must keep in mind that
this fraction may be somewhat higher, since the nebula appears to have
decelerated to 3/4 of its initial ejection speed.  In any case, the
ratio of kinetic to luminous energy associated with P~Cygni's eruption
($\la$1\%) is still far less than the eruption of $\eta$ Carinae in
the 1840's, when KE/Lt$\ga$1 (Smith et al.\ 2003).

While the total kinetic energy associated with P Cygni's 17th century
eruption is about 4000 times less than the 1840's eruption of $\eta$
Carinae, it comes much closer to the mechanical energy of $\eta$ Car's
second eruption in 1890, which was about 8$\times$10$^{46}$ ergs
(Smith 2005).  The similar amounts of kinetic energy in these two
outbursts is underscored by the similar properties in the two nebulae.
P~Cygni's shell and the Little Homunculus of $\eta$ Car (produced in
that 1890 outburst; Ishibashi et al.\ 2003; Smith 2005) both have
extremely bright [Fe~{\sc ii}] emission (Smith 2002a; 2005), both have
little or no detectable thermal-IR emission from dust (Smith et al.\
2003; Waters \& Wesselius 1986), they have similar amounts of mass,
and both have similar expansion velocities slower than the present-day
stellar wind that is overtaking them.

\subsection{Broader Implications for Mass-loss}

The mass and kinetic energy of P~Cygni's nebula are important for
understanding the mechanism behind giant eruptions of LBVs and
extragalactic $\eta$~Car analogs, as well as their role in mass loss
during the late evolution of the most massive stars.

P Cygni and $\eta$ Car are the only two LBVs in our Galaxy for which a
giant outburst has been observed directly.  Compared to $\eta$~Car
(Smith et al.\ 2003), the mass and kinetic energy of P Cygni's nebula
are much lower, and so these two objects define a range of likely
values for the mass and mechanical energy of extragalactic $\eta$~Car
eruptions.  The low mass and relatively low dust content of the nebula
around P Cygni are probably more representative of lower-luminosity
events like SN1954J (V12 in NGC~2403; Tamann \& Sandage 1968; Smith et
al.\ 2001b; van Dyk et al.\ 2005), while the massive and dusty
Homunculus around $\eta$~Car itself is more akin to the most extreme
events like SN1961V in NGC1058 (van Dyk et al.\ 2002; Filippenko et
al.\ 1995; Goodrich et al.\ 1989).

This study raises an important question in the evolution of very
massive stars: {\it As massive stars shed their outer layers to become
WR stars, is most of the mass lost in a steady stellar wind or in
brief eruptions?}  In the case of $\eta$ Car, the huge mass of the
Homunculus ($\ga$10~M$_{\odot}$) indicates that most
post-main-sequence mass loss happens during brief outbursts (Smith et
al.\ 2003).  Is this also true for P Cygni?  The total mass inside the
nebular shell of P Cygni, filled by the stellar wind since the 1600
eruption, is only 0.012 M$_{\odot}$ (assuming a steady mass loss rate
of $\dot{M}$=3$\times$10$^{-5}$ M$_{\odot}$ yr$^{-1}$; Najarro et al.\
1997).  Thus, P Cygni has lost about 10 times more mass during a brief
outburst than it has in the time since that eruption --- but obviously
this depends on the typical time period between such eruptions, which
could be longer than the 400 years since P Cygni's last outburst.
Meaburn et al.\ (1996) have studied an outer shell with a dynamical
age of roughly 2100 yr, presumably containing a similar amount of mass
to that of the inner shell.  The mass lost since that time by the
quiescent stellar wind is only about 0.07 M$_{\odot}$, supporting the
idea that most mass loss occurs during major outbursts.

Since giant outbursts like those of P Cygni, $\eta$ Car, and the
extragalactic $\eta$ Car analogs may dominate the total mass lost on
the way toward the WR phase, it is more than a little embarrassing for
the study of stellar evolution that we do not yet know how these
eruptions are triggered or what supplies their energy.

\section{SUMMARY AND CONCLUSIONS}

Our study of the near-IR spectrum of P Cygni's nebula has led to the
main conclusions below:

1.  The near-IR spectrum of P Cygni is dominated by
    collisionally-excited lines of [Fe~{\sc ii}], mainly from the
    three lowest terms.  It closely resembles the near-IR spectrum of
    the Homunculus nebula around $\eta$~Car, except that P Cygni's
    nebula lacks the molecular hydrogen seen there (Smith 2002b).

2.  Using 17 near-IR lines of [Fe~{\sc ii}] we have determined
    empirical values for the transition ratios and Einstein A
    coefficients for lines between the lowest 3 terms of Fe$^+$
    ($^6D$, $^4F$, and $^4D$).  These empirical values are in better
    agreement with the calculations of NS88 than with those of Q96,
    but in some cases our values differ substantially from both.

3.  The ratio $\lambda$12567/$\lambda$16435 is of particular
    importance, as the two bright lines share a common upper level and
    are often used to deduce the reddening.  We find an
    experimentally-determined value for this intrinsic line ratio of
    1.49 (adopting $A_V$=1.89; Lamers et al.\ 1983); higher than
    previous values.

4.  From high-resolution spectra of [Fe~{\sc ii}], the geometric
    parameters of P Cygni's inner shell are as follows: The shell
    radius is 10$^{17.3}$--10$^{17.4}$ cm, the average expansion
    velocity is 136 km s$^{-1}$, the dynamic age is about 530 yr, and
    the geometric filling factor for [Fe~{\sc ii}]-emitting filaments
    appears to be roughly 0.2.

5.  Based on the present expansion speed compared to the speed of the
    stellar wind, the dynamic age of the nebula, and the lack of any
    bright nebulosity inside the main shell, we conclude that the
    10\arcsec-radius [Fe~{\sc ii}]-emitting shell around P Cygni is
    the major product of the 1600 A.D. outburst.

6.  The electron density in the shell ranges from roughly
    2,000--11,000 cm$^{-3}$, with a representative value of 6,000
    cm$^{-3}$ for the denser [Fe~{\sc ii}]-emitting filaments.

7.  Based on models for the observed [N~{\sc ii}]
    $\lambda$6583/[N~{\sc i}] $\lambda$5200 line ratio, we find a
    likely hydrogen ionization fraction of 0.8--0.95 for a wide range
    of possible excitation scenarios, while He is mostly neutral.

8.  The total mass of the nebula is about 0.1~$M_{\odot}$, subject to
    errors in the filling factor and the electron density.  This,
    combined with the weak thermal-IR emission from dust, suggests
    that the gas:dust mass ratio exceeds the canonical value of 100 by
    an order of magnitude. Destruction of dust grains that released Fe
    into the gas phase may help explain the very bright [Fe~{\sc ii}]
    emission in P Cygni's nebula.

9.  From the total mass and average expansion velocity, the kinetic
    energy imparted to the ejecta during the 1600 A.D. outburst was
    roughly 10$^{46.3}$ ergs.

\acknowledgments \scriptsize

We are grateful to Richard Green for granting a half-night of
director's discretionary time to obtain optical spectra of P Cygni's
nebula at KPNO, and to Robert D.\ Gehrz for assistance during the
IRTF/Spex observing run. We thank Paco Najarro and John Hillier for
sharing results prior to publication regarding their recent work on
radiative transfer modeling of P Cygni's atmosphere and wind.  N.S.\
was supported by NASA through grant HF-01166.01A from the Space
Telescope Science Institute, which is operated by the Association of
Universities for Research in Astronomy, Inc., under NASA contract
NAS5-26555.

% REFERENCES

% TABLE 1 --- 
\begin{deluxetable}{llcc}
%\tabletypesize{\normalsize}
\tabletypesize{\scriptsize}
\tablecaption{Observed line intensities in the Spex spectrum}
\tablewidth{0pt}
\tablehead{
\colhead{$\lambda$(Obs.)} &\colhead{ID}  &\colhead{Intensity} &\colhead{Error} \\ 
\colhead{($\micron$)} &\colhead{\ }&\colhead{(erg s$^{-1}$ cm$^{-2}$)} 
   &\colhead{($\pm$ erg s$^{-1}$ cm$^{-2}$)}
}
\startdata

1.03199 &He~{\sc i}			&3.4$\times$10$^{-21}$	&1.0$\times$10$^{-21}$ \\
1.03931 &[N~{\sc i}]  			&4.9$\times$10$^{-21}$	&1.5$\times$10$^{-21}$ \\
1.04612 &[Ni~{\sc ii}]    		&4.4$\times$10$^{-21}$	&2.0$\times$10$^{-21}$ \\
1.05428 &N~{\sc i}    			&7.4$\times$10$^{-21}$	&1.9$\times$10$^{-21}$ \\
1.07112 &[Ni~{\sc ii}]    		&4.2$\times$10$^{-21}$	&1.4$\times$10$^{-21}$ \\
1.08277 &He~{\sc i}    			&2.17$\times$10$^{-20}$	&1.3$\times$10$^{-21}$ \\
1.09123 &He~{\sc i}+[Ni~{\sc ii}]	&5.7$\times$10$^{-21}$	&1.3$\times$10$^{-21}$ \\
1.09392 &Pa$\gamma$   			&1.54$\times$10$^{-20}$	&2.0$\times$10$^{-21}$ \\
1.10603 &?    				&6.4$\times$10$^{-21}$	&1.7$\times$10$^{-21}$ \\
1.11644 &?    				&4.9$\times$10$^{-21}$	&1.5$\times$10$^{-21}$ \\
1.12262 &?    				&4.7$\times$10$^{-21}$	&1.5$\times$10$^{-21}$ \\
1.12890 &O~{\sc i}     			&5.9$\times$10$^{-21}$	&1.7$\times$10$^{-21}$ \\
1.13477 &[Fe~{\sc ii}]			&1.38$\times$10$^{-20}$	&2.0$\times$10$^{-21}$ \\
1.23249 &N~{\sc i}    			&2.9$\times$10$^{-21}$	&0.5$\times$10$^{-21}$ \\
1.24819 &[Fe~{\sc ii}]    			&4.7$\times$10$^{-21}$	&0.8$\times$10$^{-21}$ \\
1.25642 &[Fe~{\sc ii}] ($a^6D_9-a^4D_7$)   	&1.904$\times$10$^{-19}$&0.8$\times$10$^{-21}$ \\
1.27005 &[Fe~{\sc ii}] ($a^6D_1-a^4D_1$)   	&5.4$\times$10$^{-21}$	&0.5$\times$10$^{-21}$ \\
1.27855 &[Fe~{\sc ii}] ($a^6D_3-a^4D_3$)   	&1.12$\times$10$^{-20}$	&0.9$\times$10$^{-21}$ \\
1.28177 &Pa$\beta$    				&2.58$\times$10$^{-20}$	&1.0$\times$10$^{-21}$ \\
1.28521 &Fe~{\sc ii}				&2.1$\times$10$^{-21}$	&0.4$\times$10$^{-21}$ \\
1.29396 &[Fe~{\sc ii}] ($a^6D_5-a^4D_5$)   	&2.64$\times$10$^{-20}$	&0.6$\times$10$^{-21}$ \\
1.29749 &[Fe~{\sc ii}] ($a^6D_1-a^4D_3$)    	&4.2$\times$10$^{-21}$	&0.7$\times$10$^{-21}$ \\
1.30014 &[Fe~{\sc ii}]				&1.4$\times$10$^{-21}$	&0.4$\times$10$^{-21}$ \\
1.31615 &O~{\sc i}    				&1.51$\times$10$^{-20}$	&1.0$\times$10$^{-21}$ \\
1.32028 &[Fe~{\sc ii}] ($a^6D_7-a^4D_7$)    	&5.27$\times$10$^{-20}$	&1.4$\times$10$^{-21}$ \\
1.32746 &[Fe~{\sc ii}] ($a^6D_3-a^4D_5$)    	&1.78$\times$10$^{-20}$	&1.0$\times$10$^{-21}$ \\

1.53309 &[Fe~{\sc ii}] ($a^4F_9-a^4D_5$)    	&2.23$\times$10$^{-20}$	&0.2$\times$10$^{-21}$ \\
1.59905 &[Fe~{\sc ii}] ($a^4F_7-a^4D_3$)    	&8.6$\times$10$^{-21}$	&0.3$\times$10$^{-21}$ \\
1.63735 &He~{\sc i}  				&1.3$\times$10$^{-21}$	&0.3$\times$10$^{-21}$ \\
1.64319 &[Fe~{\sc ii}] ($a^4F_9-a^4D_7$)    	&1.519$\times$10$^{-19}$&0.9$\times$10$^{-21}$ \\
1.65759 &He~{\sc i}				&2.6$\times$10$^{-21}$	&0.5$\times$10$^{-21}$ \\
1.66369 &[Fe~{\sc ii}] ($a^4F_5-a^4D_1$)    	&4.5$\times$10$^{-21}$	&0.5$\times$10$^{-21}$ \\
1.67664 &[Fe~{\sc ii}] ($a^4F_7-a^4D_5$)    	&2.00$\times$10$^{-20}$	&0.6$\times$10$^{-21}$ \\
1.68121 &Br11  					&2.7$\times$10$^{-21}$	&0.5$\times$10$^{-21}$ \\
1.68713 &Fe~{\sc ii}    			&5.9$\times$10$^{-21}$	&0.5$\times$10$^{-21}$ \\
1.69731 &He~{\sc i}  				&2.2$\times$10$^{-21}$	&0.3$\times$10$^{-21}$ \\
1.70099 &He~{\sc i}    				&4.0$\times$10$^{-21}$	&0.5$\times$10$^{-21}$ \\
1.71101 &[Fe~{\sc ii}] ($a^4F_5-a^4D_3$)    	&2.7$\times$10$^{-21}$	&0.3$\times$10$^{-21}$ \\
1.73674 &Br10				    	&4.2$\times$10$^{-21}$	&0.5$\times$10$^{-21}$ \\
1.74497 &[Fe~{\sc ii}] ($a^4F_3-a^4D_1$)    	&3.9$\times$10$^{-21}$	&0.6$\times$10$^{-21}$ \\
1.79736 &[Fe~{\sc ii}] ($a^4F_3-a^4D_3$)    	&3.9$\times$10$^{-21}$	&0.9$\times$10$^{-21}$ \\
1.80033 &[Fe~{\sc ii}] ($a^4F_5-a^4D_5$)    	&9.6$\times$10$^{-21}$	&0.8$\times$10$^{-21}$ \\
1.80963 &[Fe~{\sc ii}] ($a^4F_7-a^4D_7$)    	&3.62$\times$10$^{-20}$	&1.2$\times$10$^{-21}$ \\

2.05835 &He~{\sc i}    			&7.3$\times$10$^{-21}$	&1.0$\times$10$^{-21}$ \\
2.14273 &Mg~{\sc ii} (?)		&3.6$\times$10$^{-21}$	&0.7$\times$10$^{-21}$ \\
2.16547 &Br$\gamma$    			&8.2$\times$10$^{-21}$	&1.1$\times$10$^{-21}$ \\
2.22333 &[Fe~{\sc ii}]    		&1.8$\times$10$^{-21}$	&0.6$\times$10$^{-21}$ \\
2.30786 &[Ni~{\sc ii}]	($a^4F-a^2F$)	&2.7$\times$10$^{-21}$	&0.7$\times$10$^{-21}$ \\
2.36907 &[Ni~{\sc ii}]	($a^4F-a^2F$)	&2.3$\times$10$^{-21}$	&0.7$\times$10$^{-21}$ \\
2.40651 &?    				&4.1$\times$10$^{-21}$	&1.2$\times$10$^{-21}$ \\

\enddata
\end{deluxetable}

% TABLE 2 --- 
\begin{deluxetable}{lccccccc}
%\tabletypesize{\normalsize}
\tabletypesize{\scriptsize}
\tablecaption{Ratios of Einstein A Values for Near-IR [Fe~{\sc ii}]
  Emission Lines}
\tablewidth{0pt}
\tablehead{
\colhead{\ } &\colhead{$A_V$=0} &\colhead{$A_V$=0} &\colhead{$A_V$=1.89}
&\colhead{$A_V$=1.89} &\colhead{$A_V$=1.89} &\colhead{\ } &\colhead{\ } \\
\colhead{Transition Ratio} &\colhead{Q} &\colhead{NS} &\colhead{Q}
&\colhead{NS} &\colhead{New} &\colhead{Obs.} &\colhead{Err.}
}
\startdata

10-1/10-2  1.25668/1.32055   &3.80  &3.82   &3.66  &3.68  &3.62   &3.61 &0.10  \\
10-1/10-6  1.25668/1.64355   &1.04  &1.36   &0.87  &1.15  &1.25   &1.25 &0.01  \\
10-1/10-7  1.25668/1.80939   &5.17  &6.75   &4.17  &5.44  &5.24   &5.25 &0.18  \\
10-2/10-6  1.32055/1.64355   &0.27  &0.36   &0.24  &0.31  &0.35   &0.35 &0.01  \\
10-2/10-7  1.32055/1.80939   &1.36  &1.77   &1.14  &1.48  &1.45   &1.46 &0.06  \\
10-6/10-7  1.64355/1.80939   &4.99  &4.97   &4.77  &4.75  &4.19   &4.20 &0.14  \\
11-2/11-3  1.24819/1.29427&\nodata&\nodata&\nodata&\nodata&0.35   &0.35 &0.06  \\
11-3/11-4  1.29427/1.32778   &1.74  &1.64   &1.70  &1.61  &1.48   &1.48 &0.09  \\
11-3/11-6  1.29427/1.53347   &0.75  &0.94   &0.67  &0.84  &1.18   &1.18 &0.03  \\
11-3/11-7  1.29427/1.67688   &1.03  &1.30   &0.88  &1.11  &1.32   &1.32 &0.05  \\
11-3/11-8  1.29427/1.80002   &1.51  &1.89   &1.25  &1.56  &2.75   &2.75 &0.24  \\
11-4/11-6  1.32778/1.53347   &0.43  &0.57   &0.39  &0.52  &0.80   &0.80 &0.05  \\
11-4/11-7  1.32778/1.67688   &0.59  &0.79   &0.52  &0.69  &0.89   &0.89 &0.06  \\
11-4/11-8  1.32778/1.80002   &0.87  &1.15   &0.73  &0.97  &1.86   &1.85 &0.19  \\
11-6/11-7  1.53347/1.67688   &1.37  &1.38   &1.31  &1.31  &1.11   &1.12 &0.03  \\
11-6/11-8  1.53347/1.80002   &2.01  &2.00   &1.86  &1.85  &2.33   &2.32 &0.19  \\
11-7/11-8  1.67688/1.80002   &1.47  &1.46   &1.42  &1.41  &2.09   &2.08 &0.18  \\
12-4/12-5  1.27878/1.29777   &2.30  &2.28   &2.28  &2.26  &2.67   &2.67 &0.49  \\
12-4/12-7  1.27878/1.59947   &0.73  &0.86   &0.64  &0.74  &1.31   &1.30 &0.11  \\
12-4/12-8  1.27878/1.71113   &2.78  &3.27   &2.33  &2.74  &4.14   &4.15 &0.57  \\
12-4/12-9  1.27878/1.79710   &1.62  &1.89   &1.33  &1.55  &2.89   &2.87 &0.70  \\
12-5/12-7  1.29777/1.59947   &0.32  &0.38   &0.28  &0.33  &0.49   &0.49 &0.08  \\
12-5/12-8  1.29777/1.71113   &1.21  &1.43   &1.02  &1.21  &1.55   &1.56 &0.31  \\
12-5/12-9  1.29777/1.79710   &0.71  &0.83   &0.58  &0.69  &1.08   &1.08 &0.31  \\
12-7/12-8  1.59947/1.71113   &3.79  &3.81   &3.66  &3.69  &3.16   &3.19 &0.37  \\
12-7/12-9  1.59947/1.79710   &2.22  &2.21   &2.09  &2.09  &2.20   &2.21 &0.51  \\
12-8/12-9  1.71113/1.79710   &0.58  &0.58   &0.57  &0.57  &0.70   &0.69 &0.18  \\
13-5/13-8  1.27035/1.66377   &0.92  &1.02   &0.77  &0.86  &1.20   &1.20 &0.17  \\
13-5/13-9  1.27035/1.74493   &1.85  &2.05   &1.52  &1.69  &1.38   &1.38 &0.25  \\
13-8/13-9  1.66377/1.74493   &2.02  &2.01   &1.97  &1.96  &1.15   &1.15 &0.22  \\
\enddata
%\tablenotetext{a}{}
\tablecomments{``Q'' refers to Quinet et al.\ (1996), ``NS'' refers to
  Nussbaumer \& Storey (1988), and ``New'' refers to empirical values
  determined in this work.  Wavelengths in air are taken from Quinet
  et al.\ (1996).}
\end{deluxetable}

% TABLE 3 --- 
\begin{deluxetable}{lccc}
%\tabletypesize{\normalsize}
\tabletypesize{\scriptsize}
\tablecaption{Einstein A Values}
\tablewidth{0pt}
\tablehead{
\colhead{Transition} &\colhead{$A_{\rm Q}$} &\colhead{$A_{\rm NS}$} &\colhead{$A_{\rm New}$}
}
\startdata
10-1 12566.80  &4.74$\times$10$^{-3}$  &4.83$\times$10$^{-3}$  &4.83$\times$10$^{-3}$ \\
10-2 13205.54  &1.31$\times$10$^{-3}$  &1.33$\times$10$^{-3}$  &1.35$\times$10$^{-3}$ \\
10-6 16435.50  &5.98$\times$10$^{-3}$  &4.65$\times$10$^{-3}$  &4.26$\times$10$^{-3}$ \\
10-7 18093.95  &1.32$\times$10$^{-3}$  &1.03$\times$10$^{-3}$  &1.07$\times$10$^{-3}$ \\
11-2 12481.9   &\nodata                &\nodata                &0.35$\times$10$^{-3}$ \\
11-3 12942.68  &1.98$\times$10$^{-3}$  &1.94$\times$10$^{-3}$  &1.94$\times$10$^{-3}$ \\
11-4 13277.76  &1.17$\times$10$^{-3}$  &1.21$\times$10$^{-3}$  &1.32$\times$10$^{-3}$ \\
11-6 15334.71  &3.12$\times$10$^{-3}$  &2.44$\times$10$^{-3}$  &1.74$\times$10$^{-3}$ \\
11-7 16768.76  &2.49$\times$10$^{-3}$  &1.94$\times$10$^{-3}$  &1.63$\times$10$^{-3}$ \\
11-8 18000.16  &1.82$\times$10$^{-3}$  &1.43$\times$10$^{-3}$  &0.81$\times$10$^{-3}$ \\
12-4 12787.76  &2.45$\times$10$^{-3}$  &2.25$\times$10$^{-3}$  &2.66$\times$10$^{-3}$ \\
12-5 12977.73  &1.08$\times$10$^{-3}$  &1.00$\times$10$^{-3}$  &1.00$\times$10$^{-3}$ \\
12-7 15994.73  &4.18$\times$10$^{-3}$  &3.28$\times$10$^{-3}$  &2.20$\times$10$^{-3}$ \\
12-8 17111.29  &1.18$\times$10$^{-3}$  &0.92$\times$10$^{-3}$  &0.72$\times$10$^{-3}$ \\
12-9 17971.04  &2.12$\times$10$^{-3}$  &1.67$\times$10$^{-3}$  &1.06$\times$10$^{-3}$ \\
13-5 12703.46  &3.32$\times$10$^{-3}$  &2.91$\times$10$^{-3}$  &2.91$\times$10$^{-3}$ \\
13-8 16637.66  &4.75$\times$10$^{-3}$  &3.73$\times$10$^{-3}$  &2.68$\times$10$^{-3}$ \\
13-9 17449.34  &2.47$\times$10$^{-3}$  &1.95$\times$10$^{-3}$  &2.39$\times$10$^{-3}$ \\
\enddata
%\tablenotetext{a}{}
\tablecomments{``Q'' refers to Quinet et al.\ (1996), ``NS'' refers to
  Nussbaumer \& Storey (1988), and ``New'' refers to empirical values
  determined in this work.  Wavelengths in air are taken from Quinet
  et al.\ (1996).}
\end{deluxetable}

% TABLE 4 --- 
\begin{deluxetable}{llccc}
%\tabletypesize{\normalsize}
\tabletypesize{\scriptsize}
\tablecaption{Adopted Shell Parameters}
\tablewidth{0pt}
\tablehead{
\colhead{Parameter} &\colhead{Units} &\colhead{Inner} 
  &\colhead{Outer} &\colhead{Total}
}
\startdata
v$_{\rm sys}$ (heliocentric)	&km s$^{-1}$	&\nodata&\nodata&--38	\\
Radial expansion velocity	&km s$^{-1}$	&120	&152	&136	\\
Radius ($R_1$ and $R_2$)	&arcseconds	&7.8	&9.7	&8.8	\\
Radius ($R_1$ and $R_2$) 	&log$_{10}$ cm	&17.3	&17.4	&17.35	\\
Dynamic age (R/V)		&years		&531	&524	&527	\\
Filling factor			&\nodata	&\nodata&\nodata&0.2	\\
Avg.\ electron density		&cm$^{-3}$	&\nodata&\nodata&6000	\\
H ionization fraction		&\nodata	&\nodata&\nodata&0.8-0.95 \\
Mass				&M$_{\odot}$    &\nodata&\nodata&0.1	\\
Kinetic energy			&log$_{10}$ ergs&\nodata&\nodata&46.3	\\
KE/Lt				&\nodata        &\nodata&\nodata&$\la$0.01 \\

\enddata
%\tablenotetext{a}{}
\end{deluxetable}

% FIGURE 1 ---------- 
\begin{figure}
\epsscale{0.5}
\plotone{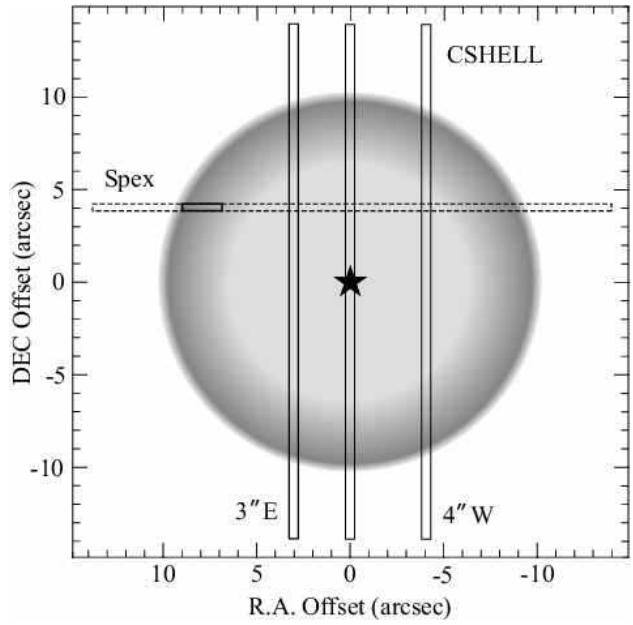}
\caption{Long-slit CSHELL and Spex aperture positions and orientations
superposed on an idealized representation of P Cygni's inner shell
nebula.  The small solid segment of the dashed Spex slit is the
extraction region for the spectra in Figures 2 and 3.  The vertical
CSHELL slits correspond to the 2-D position-velocity diagams in Figure
5.  The slit position used for the visual-wavelength Goldcam spectra
from KPNO was identical to the Spex slit position.}
\end{figure}

% FIGURE 2 -----------
\begin{figure}
\epsscale{0.6}
\plotone{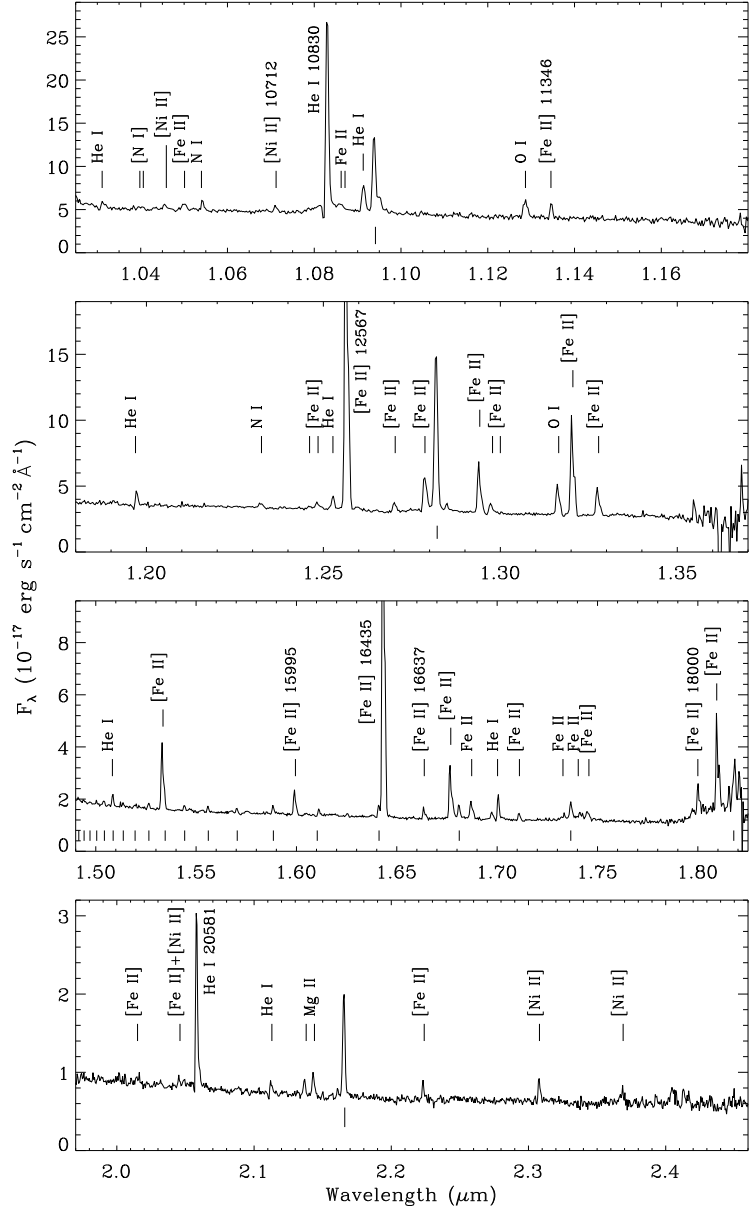}
\caption{Near-IR spectrum of P Cyg's nebula observed with Spex,
extracted from a 2\arcsec\ wide segment of the slit, centered
4\arcsec\ north and 8\arcsec\ east of the star.  The continuum and
some of the line emission is due to scattered starlight.  Hydrogen
lines are marked with un-labeled dashes.}
\end{figure}

% FIGURE 3 ------------
\begin{figure}
\epsscale{0.6}
\plotone{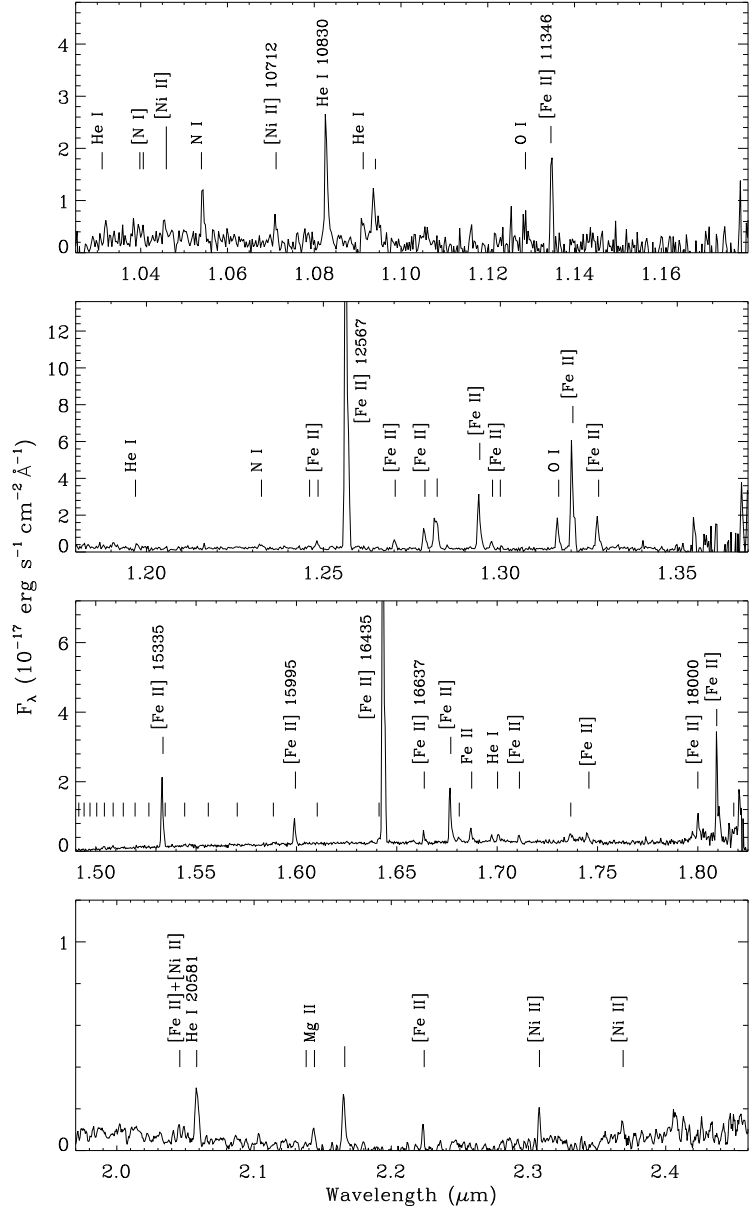}
\caption{Same as Figure 2, but with scattered starlight subtracted.
The scattered light spectrum was sampled from a region closer to the
star, where the reflected light was much stronger.  The residual
emission is intrinsic nebular emission from the shell around P Cygni.}
\end{figure}

% FIGURE 4 ------------
\begin{figure}
\epsscale{0.4}
\plotone{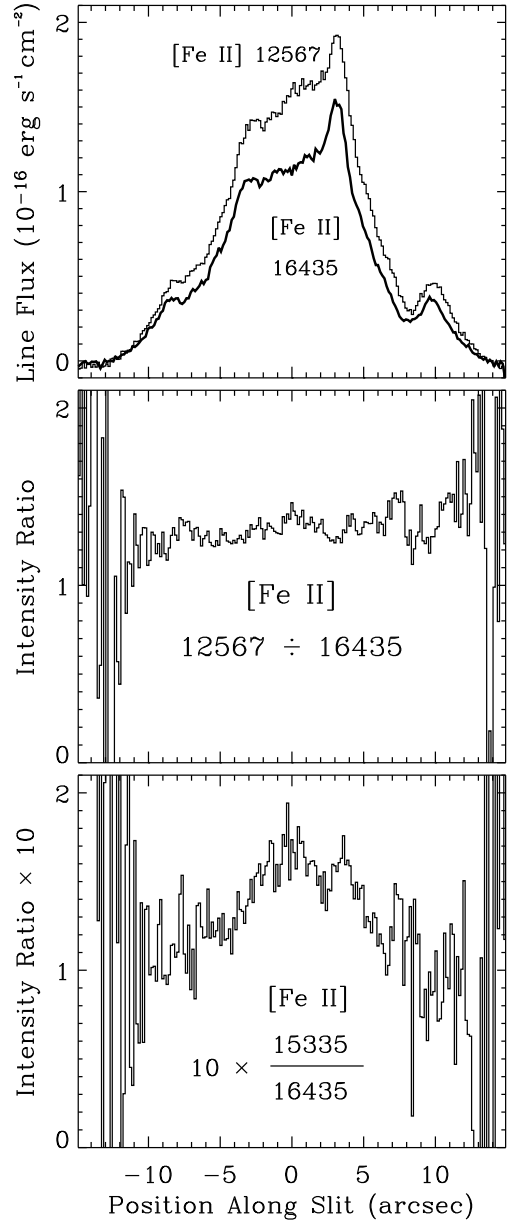}
\caption{Tracings along the long-slit Spex aperture.  Top:  Intensity
  of [Fe~{\sc ii}] $\lambda$12567 and $\lambda$16435.   Middle:  The
  [Fe~{\sc ii}] $\lambda$12567/$\lambda$16435 flux ratio.  Bottom:
  The [Fe~{\sc ii}] $\lambda$15335/$\lambda$16435 flux ratio, which
  is a potential diagnostic of electron density.}
\end{figure}

% FIGURE 5 ------------
\begin{figure}
\epsscale{0.5}
\plotone{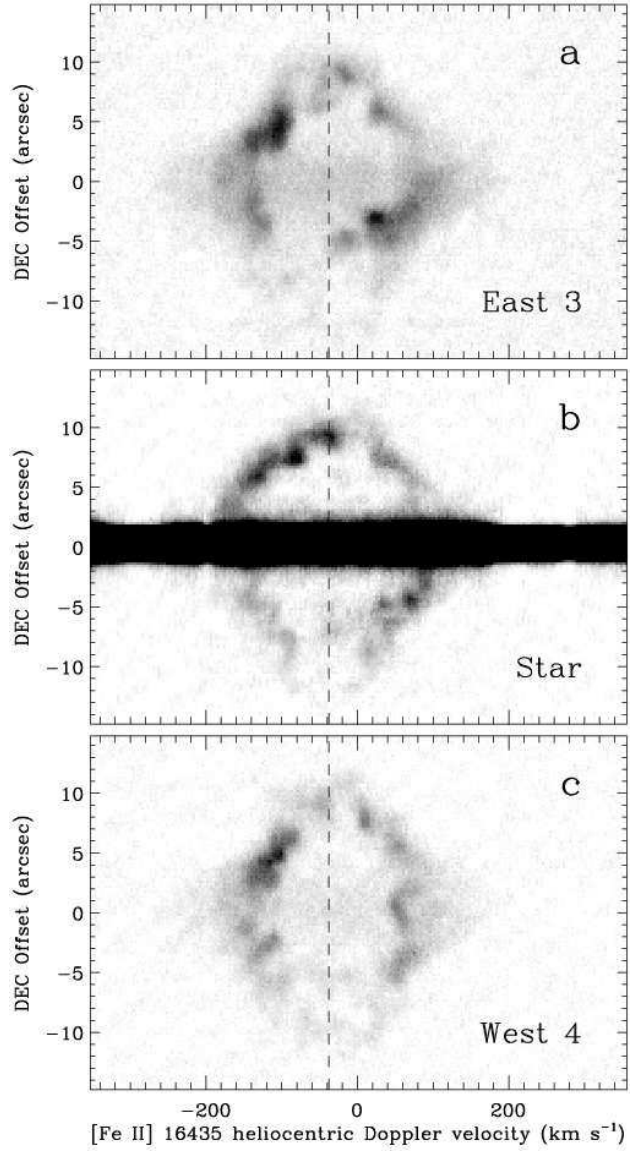}
\caption{Long-slit spectra of P Cygni obtained with CSHELL, showing
  the kinematics of the nebula.  Spectra were obtained with the slit
  oriented at P.A.=0\arcdeg\ and positioned (a) 3\arcsec\ east of the
  star, (b) centered on the star, and (c) 4\arcsec\ west of the star.
  Heliocentric velocities are shown here, and in each panel, the
  dashed vertical line marks the presumed systemic velocity of $-$38
  km s$^{-1}$.}
\end{figure}

% FIGURE 6 ------------
\begin{figure}
\epsscale{0.5}
\plotone{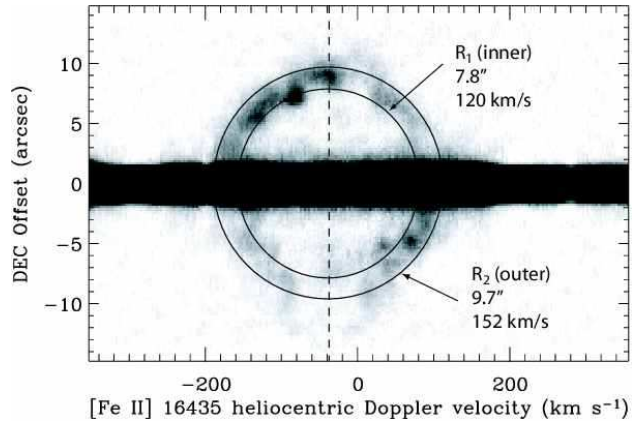}
\caption{Same as Figure 5$b$ for the slit passing through the star,
  but with two ellipses drawn to represent the inner and outer
  boundaries of a representative spherical shell (with radii of $R_1$
  and $R_2$, respectively).}
\end{figure}

% FIGURE 7 ------------
\begin{figure}
\epsscale{0.4}
\plotone{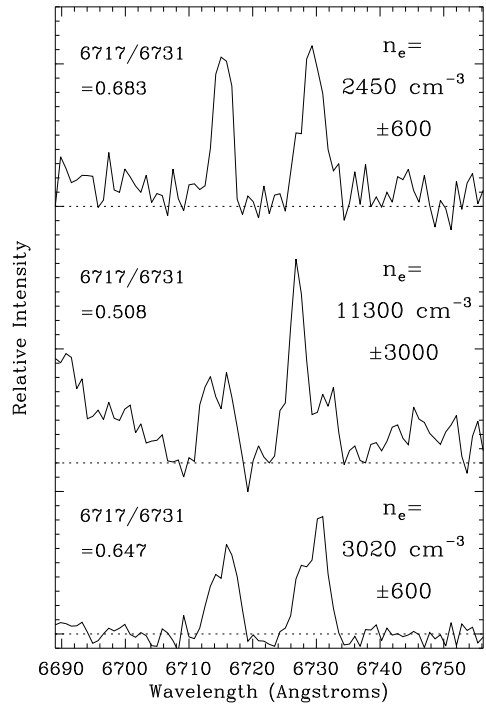}
\caption{Spectra of [S~{\sc ii}] $\lambda\lambda$6717,6731 at three
  positions across the long-slit aperture, which was positioned
  4\arcsec\ north of the star (same as the Spex slit shown in Figure
  1).  The three spectra are centered at roughly 7\arcsec\ east (top),
  the center (middle), and about 7\arcsec\ west (bottom) relative to
  the central star (see Fig.\ 1).  Measured line ratios and electron
  densities for each position are also indicated.}
\end{figure}

\end{document}